\documentclass[12pt]{article}

\usepackage[letterpaper, margin=1in]{geometry}

\usepackage{setspace} 
\usepackage{lineno} 

\usepackage[utf8]{inputenc}
\usepackage{hyperref} 
\usepackage{nameref}

\usepackage{amsmath,amssymb} 
\usepackage{graphicx} 
\usepackage{subcaption} 
\usepackage{parskip}
\usepackage{bm}
\usepackage{xr}

\usepackage{listings} 
\usepackage{xcolor}
\lstdefinestyle{pythonclassic}{
  language=Python,
  basicstyle=\ttfamily\small,
  keywordstyle=\color{blue},
  commentstyle=\color{gray},
  stringstyle=\color{red},
  frame=single,
  breaklines=true,
  showstringspaces=false
}

\usepackage{floatrow}
\usepackage[section]{placeins}

\usepackage{authblk}

\usepackage[english]{babel}
\usepackage{csquotes}

\usepackage[authoryear]{natbib}

\usepackage{amsmath,amssymb}

\DeclareMathOperator*{\argmax}{argmax}

\title{SeismoStats: A Python Package for Statistical Seismology}
\author[1]{Aron Mirwald$^*$}
\author[1]{Nicolas Schmid}
\author[1]{Leila Mizrahi}
\author[1]{Marta Han}
\author[1]{Alicia Rohnacher}
\author[1]{Vanille A. Ritz}
\author[1]{Stefan Wiemer}
\affil[1]{Swiss Seismological Service, ETH Zurich, Zurich, Switzerland}
\affil[*]{Email: aron.mirwald@sed.ethz.ch}
\affil[ ]{\textit{Published at SRL https://doi.org/10.1785/0220250266}}

\date{}
\begin{document}

\maketitle

\begin{abstract}
We introduce SeismoStats, a Python package that enables essential statistical seismology analyses. The package provides user-friendly tools to download and manipulate earthquake catalogs, to visualize them, and to estimate the a- and b-value of the Gutenberg-Richter law, or the magnitude of completeness. This comprehensively tested, well-documented, and openly accessible Python package is intended to serve as the foundation of a long-term community effort, continually expanding in functionality through shared contributions. We invite seismologists and developers to contribute ideas and code to support and shape its future development.
\end{abstract}

\section{Introduction}
\sloppy

Earthquake catalogs are a fundamental resource for seismological research and can be subjected to a wide range of analyses to understand the properties of seismicity. Some of these analyses are performed regularly by seismologists to statistically characterize seismicity. However, these analysis methods are often implemented independently by different researchers. This not only leads to extra work across the community but also increases the chance for inconsistencies and implementation errors. 

\begin{figure}[ht!]
\begin{center}
  \includegraphics[width=0.9\textwidth]{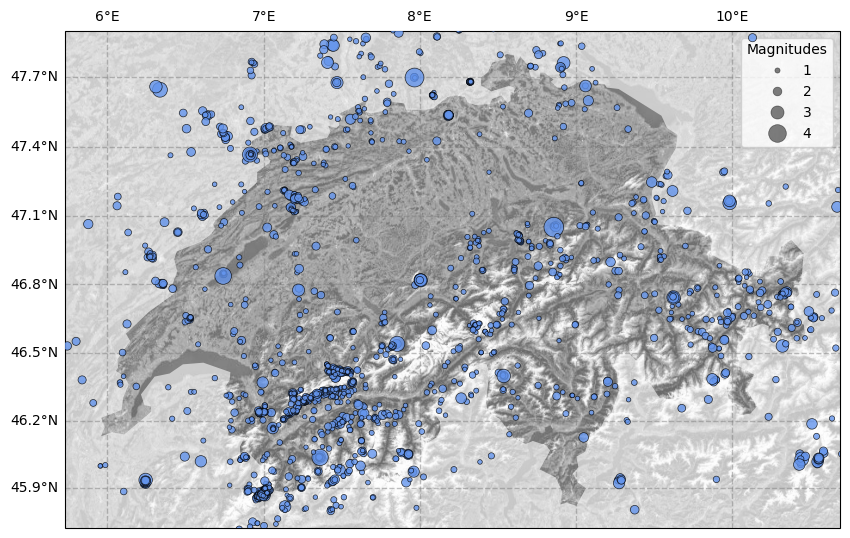}
  \caption{An example of an earthquake catalog: The Swiss earthquake catalog for the year 2024 \citep{swiss_seismological_service_national_1983}. Figure generated using the SeismoStats function \texttt{plot\_in\_space}. \\ \textit{Alt-text: Topographic map of Switzerland and the neighbouring regions in shades of grey with an overlay of the seismicity recorded in 2024 as blue dots. The dots are scaled by their magnitude. The seismicity is particularly concentrated on the Alpine arc and the Valais region alond the Rhone-Simplon axis. The largest two events occurred in eastern-central Switzerland and in southern Germany, North of the Swiss border.}}
  \label{fig:swiss_map}
  \end{center}
\end{figure}

Here, we introduce the software package SeismoStats, which aims to alleviate this. The package is designed to provide functionalities such as downloading a catalog, visualizing the seismicity and its distribution of magnitudes, and estimating statistically relevant parameters of the catalog. Figure \ref{fig:swiss_map} shows a map of seismicity in Switzerland that was created by applying one of the standard plotting functions of SeismoStats to the Swiss catalog downloaded using SeismoStats.

The main focus of SeismoStats is one of the most important relations in statistical seismology, the Gutenberg-Richter (GR) law \citep{gutenberg_frequency_1944}. Estimation and analysis of the three parameters of the GR law, magnitude of completeness ($m_c$), a-value, and b-value, have become a ubiquitous practice in statistical seismology. Many seismologists continue to use the ZMAP software for these purposes \citep{wiemer_software_2001}. However, this software is only available in the proprietary software MATLAB. In addition, it is no longer maintained, and since its last update, there have been scientific advances relevant for the estimation of the a- and b-value \citep{kijko_extension_2012, van_der_elst_bpositive_2021, van_der_elst_positive_2023, lippiello_b-more-incomplete_2024} that are not implemented in ZMAP.

SeismoStats fills this gap by providing open source tools for estimating GR parameters using the widely adopted Python programming language. The package incorporates several recent methodological improvements for estimating a- and b-values, while also offering common functionality such as downloading and visualizing earthquake catalogs, and evaluating temporal variation in the b-value. Although GR parameter estimation is currently a central feature, the package is not limited to this application. Its modular and extensible architecture supports the integration of other statistical methods, and as an open-source project under active development, SeismoStats welcomes contributions from the scientific community to ensure it continues to grow with the evolving needs of seismologists.

SeismoStats complements other Python libraries for seismology, such as ObsPy \citep{beyreuther_obspy_2010} for waveform processing, SeisComP \citep{gfz_data_services_seiscomp_2008} for real-time analysis, pyCSEP \citep{savran_pycsep_2022, graham_new_2024} for forecast testing, SeisBench \citep{woollam_seisbenchtoolbox_2022} for ML applications, and OpenQuake \citep{pagani_openquake_2014} for risk and hazard modeling. Together, these tools provide a powerful and flexible foundation for modern seismological research.

This article is structured into two parts: In the first part, we present the fundamental structure of the package, its general capabilities, and development practices. In the second part, we describe the estimation methods of GR parameters ($m_c$, b-value and a-value) that are implemented in SeismoStats.

\section{Structure and development of SeismoStats}
\begin{figure}
    \centering
    \includegraphics[width=0.8\linewidth]{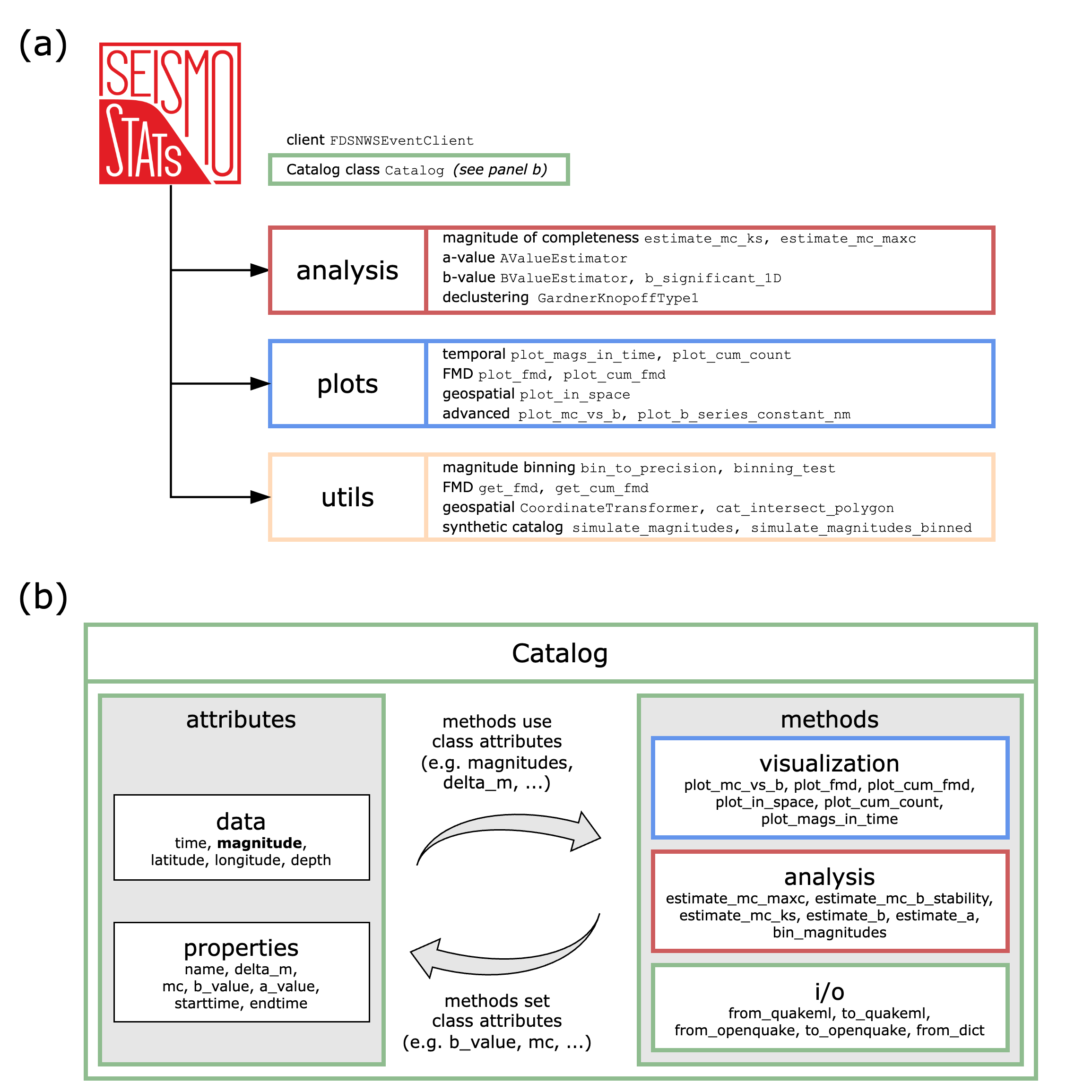}
    \caption{Overview of the SeismoStats package. a) Organisation of the subpackages with selected functions. b) Detailed architecture of the Catalog class in which the class methods can both use and update the catalog’s attributes -- properties and data.\\
    \\
  \textit{Alt-text: Two panel figure showing the  structure of SeismoStats as a schematic diagram. (a) The SeismoStats package contain a native class 'Catalog' and three main groups of functions: analysis, plots and utils. The analysis subpackage contains methods to calculate the magnitude of completeness, a-value, b-value as well as declustering algorithms. The plots subpackage contains temporal, frequency-magnitude distribution, geospatial and advanced plotting functions. The utils subpackage contains methods for magnitude binning, frequency-magnitude distribution, geospatial and sythetic catalog generation. Example of function names are provided for each type of analysis, plots and utils function. (b) This panel shows a detailed view of the structure of the Catalog class. On the left, attributes of a Catalog object are listed like data (for example time, magnitude, geographical coordinates) or properties (name, binning, magnitude of completeness, start time, end time, a-value, b-value). On the right, class methods that can be applied to a Catalog object are listed in three subcategories: visualization, analysis and input-output. For each subcategory names of class methods are listed as examples. Between the attributes and methods blocs, arrows show the interaction. On the arrow going from attributes and pointing to methods is written "methods use class attributes (for example magnitudes, binning,...). On the arrow going from methods and pointing to attributes is written "methods set class attributes (for example b-value, magnitude of completeness...).}}
    \label{fig:overview}
\end{figure}

The overarching structure of the package is twofold. First, SeismoStats simplifies access to and handling of earthquake catalogs by providing the Catalog class (see panel (b) of Figure \ref{fig:overview}). It integrates with functions to download earthquake catalogs and return them as Catalog objects, as well as conversion tools to the standard QuakeML format \citep{danijel_schorlemmer_quakeml_2011} and the Catalog format. Both the Catalog class and the download client are located at the very top level of the package.
Second, SeismoStats provides flexible, standalone functions that can operate on common Python data structures such as lists, NumPy arrays \citep{harris_array_2020}, and pandas series \citep{the_pandas_development_team_pandas-devpandas_2020}. This design allows users to incorporate the functionality of the package into their existing workflows with minimal adaptation. These functions are located within three sub-packages:  \texttt{analysis}, \texttt{utils} and \texttt{plots}.
 
\subsection{The Catalog class}
At the core of the package is the Catalog class, a subclass of the pandas DataFrame \citep{the_pandas_development_team_pandas-devpandas_2020}.

A Catalog object must contain at least a `magnitude' column, and additional columns, like `latitude', `longitude', `depth', and `time', can be added as required by the user. The initialization and manipulation of the Catalog are identical to those of a pandas DataFrame. However, the Catalog class provides additional properties and functionalities implemented as class methods, each requiring certain columns to be present in order to be available. 

Many of the functions within the sub-packages \texttt{analysis}, \texttt{utils} and \texttt{plots} are available as methods of the Catalog class, so its properties and attributes can be used and updated as needed. Examples include the estimation of the magnitude of completeness $m_c$ (which stores $m_c$ as a property) and the b-value (which uses the property $m_c$ directly), as well as visualization methods for seismicity distributions.

The Catalog class is further enhanced by the package's functionalities which facilitate the import and analysis of earthquake data. For instance, QuakeML files can be converted into a Catalog, and SeismoStats supports downloading earthquake catalogs from FDSN Web Services using the \texttt{FDSNWSEventClient}.

\subsection{Standalone functions}
The standalone functions in SeismoStats are organized into three sub-packages. \texttt{SeismoStats.analysis} provides functions for estimating and evaluating Gutenberg–Richter parameters, such as the magnitude of completeness, a-value, and b-value, as well as a method for declustering catalogs.

Estimation methods for the a-value and b-value are implemented as classes to ensure consistency across the different methods and to enhance re-usability. For example, all a- and b-value classes require the input of magnitudes, magnitude of completeness, and bin-size of the discretized magnitudes for estimation. In contrast, functions to estimate completeness are implemented as standalone functions, since they
share fewer common features. Tools to evaluate the quality and robustness of estimates include the Shi and Bolt \citep{shi_standard_1982} standard deviation of the b-value, the Lilliefors test \citep{lilliefors_kolmogorov-smirnov_1969, herrmann_inconsistencies_2021} to assess whether the Gutenberg–Richter law is a valid assumption, and the b-significance method \citep{mirwald_how_2024} to test whether the b-value varies significantly. Finally, the declustering method proposed by \cite{gardner_is_1974} is included.

\texttt{SeismoStats.plots} offers functions for visualizing seismic data and analysis results. Built on modern plotting libraries such as matplotlib, cartopy, and geopandas \citep{hunter_matplotlib_2007, met_office_cartopy_2010,the_pandas_development_team_pandas-devpandas_2020}, this module includes functions for mapping seismicity in space and time, plotting frequency-magnitude distributions (FMDs), and visualizing parameter estimates. All figures in this article are created using the plotting functionality of SeismoStats.

\texttt{SeismoStats.utils} contains utility functions that support statistical workflows. These include tools for testing the discretization of magnitudes, generating synthetic magnitudes, and preparing data for analysis, for example, by binning the magnitudes. Further capabilities are the transformation of coordinates to local ones, and filtering catalogs based on a polygon.

\subsection{Development Practices and community contributions}
SeismoStats is developed following best practices in software development. Continuous integration and continuous deployment (CI/CD) workflows are implemented to maintain stability, supported by extensive unit tests with coverage tracking that catch errors early. Semantic versioning is used to clearly manage changes and updates, and PEP 8 style guides are followed to maintain code style consistency across the package. Documentation is generated and deployed automatically through the CI/CD pipeline, which helps keep documentation up-to-date, while a user guide complements this with examples covering the basic usage.

SeismoStats leverages established Python libraries such as Cartopy \citep{met_office_cartopy_2010}, Geopandas \citep{the_pandas_development_team_pandas-devpandas_2020},  Matplotlib \citep{hunter_matplotlib_2007}, Numpy \citep{harris_array_2020}, Pandas \citep{the_pandas_development_team_pandas-devpandas_2020}, Scipy \citep{virtanen_pauli_scipy_2020}, Shapely \citep{sean_gillies_shapely_2022}, and Statsmodels \citep{seabold_statsmodels_2010}.

As an open-source project hosted on GitHub, SeismoStats is designed to be a long-lasting community effort. Any method may be implemented by members of the community, provided that it is useful for more than one person and is not excessively complex. The minimal criteria to accept a function is that it is documented reasonably, unit tested and passed a review.

\section{Implemented methods to estimate GR parameters}

The GR law states that earthquake magnitudes are exponentially distributed \citep{gutenberg_frequency_1944, ishimoto_observations_1939}.

\begin{equation}\label{eq:GR law}
    N(m) = 10^{a - b(m - m_c)},
\end{equation}
where $N(m)$ is the number of earthquakes with a magnitude equal to or larger than magnitude $m$ above a magnitude threshold $m_c$. The so-called a- and b-value are parameters that quantify the total number of earthquakes and the relative frequency of small vs. large earthquakes, respectively. The threshold $m_c$ in this context is the magnitude of completeness above which we assume to be able to detect all earthquakes.

In this section, we introduce each implemented method to estimate the magnitude of completeness, b-value, and a-value.

\subsection{Methods to evaluate completeness}

\begin{figure}[ht!]
    \centering
    \includegraphics[width=1.0\textwidth]{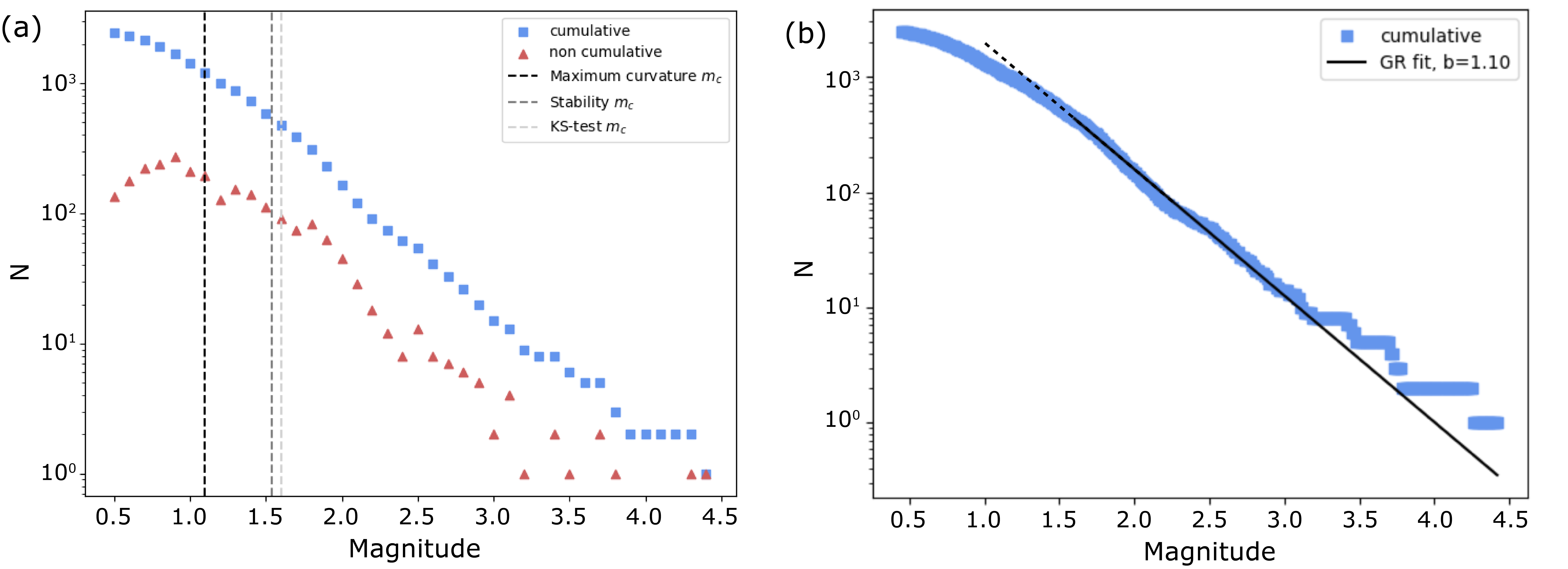}
    \caption{Visualisation of the frequency magnitude distribution of the 2024 Swiss catalog. (a) Application of the three completeness estimation methods. The blue squares represent the cumulative FMD, the red triangles the non-cumulative one. The vertical dashed lines indicate the completeness magnitude that was estimated using the different available methods. Figure generated using SeismoStats functions \texttt{plot\_fmd} and \texttt{plot\_cum\_fmd}. (b) Example of an estimated b-value fit. The estimator \texttt{ClassicBValueEstimator} was applied, with a bin size of $\Delta m = 0.01$ and assuming an incompleteness of $m_c = 1.6$. The GR-law with the estimated b-value is shown in black, the corresponding cumulative FMD is shown in blue. The dotted line is the extension of the GR to the reference magnitude M=1. Figure generated using SeismoStats function \texttt{plot\_cum\_fmd}. \\ \textit{Alt-text: Two panel figure showing the frequency-magnitude distribution of seismicity in Switzerland in 2024. (a) The incremental distribution of magnitudes with a binning of 0.1 is shown with red triangles. The cumulative distribution is shown with blue squares. Three vertical lines show the estimated magnitudes of completeness with maximum curvature (Mc=1.1), Mc-by-bvalue-stability (Mc=1.5) and Kolmogorov-Smirnov test (Mc=1.6). (b) The cumulative distribution with a binning of 0.01 is shown with blue squares. The Gutenberg-Richter law fitted to the data with a b-value of 1.1 is shown as a black line using 1.6 for the completeness. The Gutenberg-Richter fit is extended to the reference magnitude M=1.0 with a dashed line.}}
    \label{fig:FMD_Swiss}
\end{figure}

The magnitude of completeness ($m_c$) is a crucial parameter in any catalog-based statistical analysis. Without it, estimates of the a- and b-value will be adversely affected. Many methods have been proposed to determine $m_c$ \citep{woessner_assessing_2005, schorlemmer_probability_2008, ogata_analysis_1993, mignan_estimating_2012}. The SeismoStats package currently includes three commonly used approaches (shown in Figure \ref{fig:FMD_Swiss}-a). Further methods can be added by the community as they are developed.


In SeismoStats, all completeness estimation methods are implemented as functions, and also available as class methods, which return a tuple with the first entry being the estimated magnitude of completeness and the second entry provides additional outputs in a dictionary. This ensures that completeness functions can be used interchangeably despite their variable outputs. Although each function has different input parameters, the minimally required input for all is an array of magnitudes and the bin size of the discretized magnitudes, $\Delta m$ (hereafter referred to as bin size).

The implemented methods assume the magnitude of completeness to be time-independent. Although this can be a useful model in many cases, it is well known that the level of completeness can change over time \citep{helmstetter_comparison_2006, kagan_short-term_2004}. 


\subsubsection*{Maximum Curvature}
The Maximum Curvature method (MAXC) defines the magnitude of completeness of a catalog as the magnitude at which the non-cumulative frequency magnitude distribution (FMD) is maximal \citep{wiemer_minimum_2000}, plus a correction term to avoid underestimation,

\begin{equation}
    m_c = \argmax_{m_k} \left(N(m_k)\right) + \delta,
\end{equation}

where $N(m_k)$ is the number of earthquakes within a magnitude bin of width $\Delta m^*$ (hereafter called bin width) centered at $m_k$, and $\delta$ is the correction term, which is typically set to $\delta = 0.2$ \citep{woessner_assessing_2005}. 
Note that the bin width $\Delta m^*$ considered to calculate $m_c$ can differ from the bin size of the discretized magnitudes $\Delta m$. That is, a catalog can have magnitudes discretized to bins of $\Delta m=0.01$, but one might be interested in estimating $m_c$ with a precision of $\Delta m^*=0.1$. Within SeismoStats, the bin size of the discretized magnitudes $\Delta m$ is called \texttt{delta\_m}, while bin width $\Delta m^*$, which depends on the user's needs and is not a property of the data, is called \texttt{fmd\_bin}.

Implementation in SeismoStats: The function \texttt{estimate\_mc\_maxc} requires the array of magnitudes and the bin width $\Delta m^*$ (\texttt{fmd\_bin}) as input. The correction factor $\delta$ is set to 0.2 by default, as in the original article \citep{woessner_assessing_2005}, but can be specified otherwise by the user.

\subsubsection*{{$\boldsymbol{m_c}$} by b-value stability.}
Introduced by \cite{cao_temporal_2002} and refined by \cite{woessner_assessing_2005}, this method is based on the experience that the b-value for incomplete catalogs is underestimated. When excluding successively the smallest events by increasing the lower magnitude cutoff threshold, the b-value therefore increases. Under the assumption that the magnitudes are exponentially distributed, the b-value is expected to stabilize when the completeness is reached. The completeness threshold is thus defined as the magnitude at which the b-value is relatively stable in relation to its theoretical standard deviation \citep{shi_standard_1982}. In practice, this is achieved as follows:

\begin{equation}
    m_c = \text{min}\left\{ m_k \mid \text{abs} \left(\frac{1}{N} \sum_{j=1}^{N} b(m_k + j\cdot \Delta m^*) - b(m_k)\right) < \sigma_{b(m_k)} \right\},
\end{equation}
where $b(m)$ is the b-value estimate based on all magnitudes above $m$, $\sigma_b$ is its uncertainty (see Eq. \ref{eqn: shi_uncertainty}), $\Delta m^*$ is the magnitude bin size considered for $m_c$ precision and $N$ is the number of $m_c$ bins that is considered for stability. $L=N\cdot \Delta m^*$ is then the length of the magnitude range considered for stability. Note that in SeismoStats, the input required for b-value estimation is $L$, not $N$.

Implementation in SeismoStats: The function \texttt{estimate\_mc\_b\_stability} requires the array of magnitudes and the bin size of the discretized magnitudes $\Delta m$ as input. The magnitudes at which the stability should be tested, as well as the stability range can be additionally provided. By default, the function tests at all magnitudes between the minimum and maximum magnitude of the sample, and the stability range $L$ is $0.5$, as in the original definition of the method \citep{woessner_assessing_2005}.




\subsubsection*{KS-test.}
The determination of the magnitude of completeness through the Kolmogorov-Smirnov (KS) test is a statistical approach in which observed and expected cumulative distribution functions (CDFs) are compared \citep{clauset_power-law_2009, mizrahi_effect_2021}. For each candidate magnitude threshold $\hat{m}_c^k$, we test the null hypothesis that the magnitudes above the chosen threshold, $m_j \geq \hat{m}_c^k$, follow an exponential distribution. The KS-statistic, $D_{KS}^{obs}$, is defined as the maximum absolute difference between the empirical CDF, $F_n(m)$, and the theoretical CDF under the GR assumption, $P(m)$: 

\begin{equation}
    D_{KS}^{obs} = \sup_{m \geq \hat{m}_c^k} \left|F_n(m) - P(m)\right|.
\end{equation}

Because earthquake magnitudes are typically discretized and the b-value is estimated from the data, the standard Kolmogorov distribution cannot be used to determine the $p$-value directly. Instead, we approximate the significance by generating $N$ bootstrap synthetic catalogs sampled from the best-fit exponential distribution. For each synthetic catalog $j$, a synthetic distance $D_{KS}^j$ is calculated. The estimated $p$-value, $\hat{p}$, is the fraction of synthetic distances that are at least as large as the observed distance:

\begin{equation}
    \hat{p} = \frac{\#\{ D_{KS}^j \geq D_{KS}^{obs} \}}{N}, \quad \text{for } j = 1, \dots, N.
\end{equation}

Finally, the magnitude of completeness $m_c$ is defined as the lowest candidate $\hat{m}_c^k$ for which the model cannot be rejected at the significance level $\alpha$:

\begin{equation}
    m_c = \min \left\{ \hat{m}_c^k \mid \hat{p} \geq \alpha \right\}.
\end{equation}


It is important to note that KS tests are generally sensitive to the sample size. This means that with an increasingly large sample of magnitudes, increasingly minor deviations from the exponential distribution can be detected. As many real earthquake catalogs exhibit some kind of small artifacts in the magnitude distribution, this might make the KS method impractical for very large datasets.

Implementation in SeismoStats: The function \texttt{estimate\_mc\_ks} requires the array of magnitudes and the bin size of the discretized magnitudes $\Delta m$ as input.  The KS test is implemented so that $\hat{p}$ is computed through the simulation of $N$ (set to 10,000 by default) samples of discretized magnitudes. These are drawn from an exponential distribution with the b-value that was estimated from the observed sample, unless another specific b-value is given as input. The significance threshold is set to $\alpha = 0.1$ by default, which is a commonly used value. The smaller this threshold is, the stricter the KS test, resulting in a more conservative (higher) $m_c$ estimate.


\subsection{b-value Estimation Methods}
The b-value of the GR law (Eq. \ref{eq:GR law}) is a measure that quantifies the relative frequency of large vs. small earthquakes. The long-term global b-value is around one \citep{el-isa_spatiotemporal_2014}, indicating that the frequency of earthquakes increases tenfold for each unit decrease of magnitude. A lower b-value relates to a larger share of large magnitudes, while a higher b-value relates to a smaller share of large magnitudes.

\begin{figure}[ht!]
\begin{center}
  \includegraphics[width=0.9\textwidth]{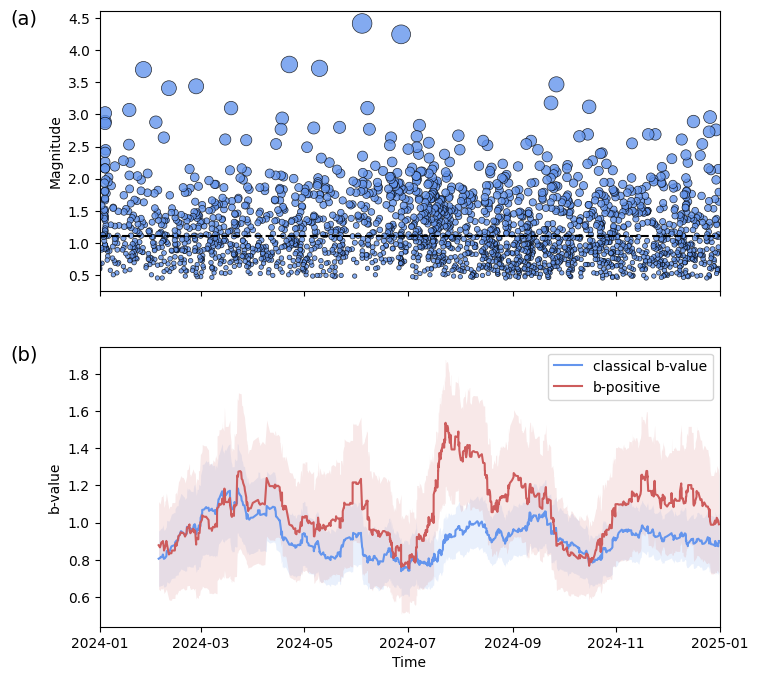}
  \caption{Illustration of temporal variation of the b-value both for the classical b-value estimate and b-positive for the 2024 Swiss catalog. (a) Scatter plot of the magnitudes over time. This figure is generated using SeismoStats function \texttt{plot\_mag\_in\_time}. (b) Classical b-value and b-positive over time, each b-value is estimated using 100 magnitudes above $m_c$. The b-value plotted at each time is estimated with magnitudes up to that time. This figure was generated using SeismoStats function \texttt{plot\_b\_series\_constant\_nm}.\\
  \textit{Alt-text: Two panel figure showing the temporal evolution of seismicity in Switzerland in 2024. (a) scatter plot of magnitudes vs. time for the 2024 Swiss catalog, with a dotted line indicating the completeness of 1.1. The largest event with a magnitude of 4.4 occurs in May 2024. (b): b-value as a function of time, estimated with the classical (in blue) and the b-positive (in red) methods. The classical b-value ranges from 0.8 to 1.2, with the highest point in March 2024. b-positive ranges from 0.8 to 1.5, with the highest point in July-August 2024. Classical and b-positive approaches are in agreement until early July 2024, from July to mid September 2024 b-positive is significantly higher that the classical b-value. In late 2024, b-positive is slightly higer than the classical b-value.}}
  \label{fig:b_seismicity_time}
  \end{center}
\end{figure}

In the case of continuous magnitudes (i.e., no discretization), the b-value can be estimated by the maximum likelihood estimation (MLE) \citep{aki_maximum_1965}:
\begin{equation}\label{eq: MLE}
    \hat{b} = \frac{\log e}{\frac{1}{n} \sum_{i=1}^n m_i - m_c},
\end{equation}
where $m_c$ is the magnitude of completeness and $m_i$ are the individual magnitudes within the catalog. All the b-value estimation methods described in this article are based on the above approach. However, they differ from it in that they consider the discretization of magnitudes, and two of the methods use magnitude differences instead of the magnitudes themselves. In Figure \ref{fig:FMD_Swiss}-b, we show an example of an estimation of the b-value. In Figure \ref{fig:b_seismicity_time}, we show an example of the b-value variation with time, estimated with two different methods.

When magnitudes are considered to be random variables, the estimate of the b-value (Eq. \ref{eq: MLE}) is also a random variable, with a mean and variance that depend on the number $n_m$ of earthquake magnitudes used for the estimate. For a sufficiently large $n_m$, the b-value estimate can be considered to follow a normal distribution \citep{shi_standard_1982}, with a standard deviation of
\begin{equation}\label{eqn: shi_uncertainty}
    \sigma(\hat{b}) = \frac{\ln(10)b^2}{\sqrt{n_m-1}}\sqrt{\text{var}(m)}.
\end{equation}
where $b$ is the true b-value, $\hat b$ is its estimate, and $\text{var}(m)$ is the variance of the magnitudes which can be estimated from the data.

All b-value methods are implemented as classes in SeismoStats. This ensures homogeneity between the implementation of different methods and has the advantage that, beside the b-value (\texttt{estimator.b\_value} or \texttt{estimator.value}), other parameters and products of the estimation can be easily accessed. These include the standard deviation of the estimate (\texttt{estimator.std}), the magnitudes above completeness used in the estimation (\texttt{estimator.magnitudes}), and the corresponding sample size (\texttt{estimator.n}). For the standard deviation, all functions use the estimate of Eq. \ref{eqn: shi_uncertainty}, with the exception of the b-more-positive method, where a bootstrap method was implemented due to the incompatibility of the former. Note that although Eq. \ref{eqn: shi_uncertainty} defines the uncertainty of the estimate assuming continuous magnitudes (Eq. \ref{eq: MLE}), in SeismoStats it is used even if the magnitudes are discretized. This is due to the lack of an alternative that does account for the discretization and the fact that it is commonly used. Finally, the Lilliefors test which tests if a given sample is exponentially distributed \citep{herrmann_inconsistencies_2021, lilliefors_kolmogorov-smirnov_1969} is implemented as a method which returns a p-value.


\subsubsection*{Classical maximum likelihood estimate} \label{section:classicalMLE}
In most catalogs of natural seismicity, the magnitudes are discretized. If not accounted for, this leads to a bias in the b-value estimate. \cite{tinti_confidence-intervals_1987} and \cite{marzocchi_review_2003} derived the exact MLE for the b-value in the case of discretized magnitudes:

\begin{equation}\label{eq: b_tinti}
    \hat{b}  =  \frac{1}{\ln(10) \Delta m} \ln p, 
\end{equation}    
where
\begin{equation}
    p = 1 + \frac{\Delta m}{\frac{1}{n} \sum_{i=1}^n m_i - m_c}.
\end{equation}

Note that for the case of continuous magnitudes ($\Delta m=0$), Eq. \ref{eq: b_tinti} is reduced to Eq. \ref{eq: MLE}.

Implementation in SeismoStats: For the estimation of the b-value, the class \texttt{ClassicBValueEstimator} requires the array of magnitudes, the magnitude of completeness and the bin size $\Delta m$ as input. Weights can be given optionally.

\subsubsection*{Utsu's approximate estimate}
As an alternative to the exact result, SeismoStats also provides an approximate introduced by \cite{utsu_method_1965}:

\begin{equation}\label{eq: MLE_utsu}
    \hat{b} = \frac{\log e}{\frac{1}{n} \sum_{i=1}^n m_i - m_c + \Delta m / 2}.
\end{equation}

This estimate, although not exact, is very commonly used and accurate for small bins.

Implementation in SeismoStats: For the estimation of the b-value, the class \texttt{UtsuBValueEstimator} requires the array of magnitudes, the magnitude of completeness, and the bin size $\Delta m$ as input.

\subsubsection*{b-positive}\label{section: b-positive}
After large earthquakes, smaller earthquakes are harder to detect, leading to a temporary drop in the fraction of small earthquakes detected. This effect is known as short-term aftershock incompleteness (STAI). STAI effectively increases the magnitude of completeness after a large event has occurred. In an effort to reduce the impact of STAI on the b-value estimate, 
\cite{van_der_elst_bpositive_2021} introduced a b-value estimation method that utilizes the positive differences of subsequent magnitudes. The method, called b-positive, effectively assumes that the completeness magnitude at each point in time is at least as large as the last earthquake magnitude recorded plus some margin ($\delta m_c$). It is based on the fact that the difference between two exponentially distributed variables follows a Laplace distribution.

The b-positive estimate is performed as follows. First, the differences between consecutive magnitudes ($m_i - m_{i-1}$) are calculated. Then, only the differences that are positive and larger than a threshold, $\text{d}m_c$, are selected (hence the name b-positive). With the obtained magnitude differences, the b-value is estimated using the classical MLE (see section \ref{section:classicalMLE}).

Although the b-positive method effectively eliminates the effect of short-term aftershock incompleteness, it is still affected by normal incompleteness resulting from the detection capabilities of the seismic network \citep{lippiello_b-more-incomplete_2024}. This effect can be reduced by considering only the complete catalog (above $m_c$) before taking the differences, or by increasing the threshold $\delta m_c$.

Implementation in SeismoStats: For the estimation of the b-value, the class \texttt{BPositiveBValueEstimator} requires the array of magnitudes, the magnitude of completeness and the bin size $\Delta m$ as input. The user can also specify the threshold $\delta m_c$ can be given as input, with the default value being one magnitude bin. If the user supplies a time array, it will be used to order the events in time, otherwise it is assumed that they are already ordered. Weights can be given optionally.

\subsubsection*{b-more-positive}\label{section:b-more-positive}
Based on the b-positive method, \cite{lippiello_b-more-incomplete_2024} developed the b-more-positive method. This method also uses magnitude differences, but the differences are constructed in an alternative way. For each earthquake of magnitude $m_i$, the next earthquake with a magnitude greater or equal than $m_i + \delta m_c$ is used to calculate the differences, effectively resulting in a higher number of magnitude differences compared to b-positive. However, this does not necessarily translate into a smaller uncertainty of the estimate. In our synthetic tests, we found that the actual standard deviation of the estimate is larger than the expected standard deviation (Eq. \ref{eqn: shi_uncertainty}). Therefore, SeismoStats uses a bootstrap method to estimate the standard deviation for the b-more-positive estimate.

Implementation in SeismoStats: For the estimation of the b-value, the class \texttt{BMorePositiveBValueEstimator} requires the array of magnitudes, the magnitude of completeness and the bin size $\Delta m$ as input. The user can also specify the threshold $\delta m_c$ can be given as input, with the default value being one magnitude bin. If the user supplies a time array, it will be used to order the events in time, otherwise it is assumed that they are already ordered. Weights can be given optionally.

\subsection{a-value Estimation Methods}
The a-value of the GR law (Eq. \ref{eq:GR law}) quantifies the number of earthquakes that occurred in a given volume and time. Together with the b-value, it is typically used to estimate the probability of an earthquake larger than $m$ \citep{tormann_systematic_2014, ritz_transient_2022}.

Similarly to the b-value estimation, the a-value estimation methods are implemented as classes in SeismoStats. One difference compared to the b-value estimator classes is that they lack the standard deviation property. Otherwise, the classes have the properties \texttt{estimator.a\_value} or \texttt{estimator.value}, \texttt{estimator.n} and \texttt{estimator.magnitudes}, similarly to the b-value estimator classes.

\subsubsection*{Classical estimate}
With Eq. \ref{eq:GR law}, we can estimate the a-value as the base-10 logarithm of the number of earthquakes above completeness,

\begin{equation}\label{eqn: a_val}
a = \log N(m_c).
\end{equation}

There are, however, two commonly used modifications of the definition above, which are relevant when comparing different a-values to each other.

First, the completeness magnitudes, relative to which different a-values are calculated, may not be constant. Therefore, in practice, the a-value is often estimated with respect to a certain reference magnitude, such that $10^{a_{m_{\text{ref}}}} = N(m_{\text{ref}})$. Here, $N(m_{\text{ref}})$ is not the observed number of earthquakes above $m_{\text{ref}}$, but the extrapolated number if the GR law were perfectly valid above and below $m_c$, as shown in Fig. \ref{fig:FMD_Swiss}-b. The new a-value can now be estimated using the a-value defined in Eq. \ref{eqn: a_val}: $a_{m_{\text{ref}}} = a - b(m_{\text{ref}} - m_c)$.

Second, the time intervals for which different a-values are compared often differ, making the comparison more difficult. In many cases, researchers scale the a-value so that $N(m)$ refers to the number of earthquakes above $m$ within a year, which is effectively a rate rather than an absolute number. This possibility is implemented in SeismoStats in the form of a scaling factor that can be provided as an input to the a-value estimator. This scaling factor quantifies how many time units fit into the observation interval. For example, if the catalog spans 10 years and the a-value is to be scaled to one year, the scaling factor is 10. Note that the scaling factor can analogously be applied to facilitate a comparison between catalogs spanning different spatial extents: To compare the number of earthquakes in two different volumes, one may be interested in the number of earthquakes per cubic km. For a volume of 100 cubic km, the scaling factor would thus be 100.

Implementation in SeismoStats: For the estimation of the a-value, the class \texttt{ClassicAValueEstimator} requires the array of magnitudes, the magnitude of completeness and the bin size $\Delta m$ as input. The user can also specify the threshold $\delta m_c$ can be given as input, with the default value being one magnitude bin. Weights can be given optionally.

\subsubsection*{a-positive}
Similarly to the b-value estimation, we have implemented `positive' a-value estimation methods. These are built on the implicit assumption that the momentary completeness magnitude is given by the magnitude of the last detected earthquake (plus a margin, called $\delta m_c$). We based our definitions on \cite{van_der_elst_positive_2023}, with the difference that we homogenized the naming convention with the b-value methods: `positive' means that the differences of magnitudes of subsequent events are used for the rate estimation, and `more positive' means that for each earthquake, the difference in magnitude to the next larger event in the catalog is used for the rate estimation. The latter method is described by \cite{van_der_elst_positive_2023}.

Both methods estimate the share of time covered by the inter-event times used to estimate the a-value. For the a-positive method, this is fairly straightforward: for each positive difference $m_{i+1} - m_i > \delta m_c$, the corresponding time difference is $\Delta t_i = t_{i+1} - t_i$. Then, the a-value can be estimated as:

\begin{equation}
    a^+ = \log n^+ - \log \frac{\sum_{i=1}^{n^+} \Delta t_i}{T},
\end{equation}

where $n^+$ is the number of positive differences in the catalog, and $T$ is the length of the entire observation interval. The result of this computation can be directly compared with the classical a-value.

Implementation in SeismoStats: For the estimation of the a-value, the class \texttt{APositiveAValueEstimator} requires the array of magnitudes, the array of times, the magnitude of completeness and the bin size $\Delta m$ as input. The user can also specify the threshold $\delta m_c$, with the default value being one magnitude bin. Weights can be given optionally.

\subsubsection*{a-more-positive} 
For the a-more-positive method (as described in \cite{van_der_elst_positive_2023}), time differences $\Delta t_i$ to the next larger event are calculated for each event. The time differences are conditioned on the magnitude being larger than the previous one, so in order to estimate the rate above $m_c$, they have to be scaled according to the GR law,
\begin{equation}
    \tau_i = \Delta t_i 10^{-b(m_i + \delta m_c)},
\end{equation}
where $m_i$ is the magnitude of the first earthquake of the pair of earthquakes with the time difference $\Delta t_i$. Therefore, this method depends on the b-value as an input. 

Next, we must take into account the earthquakes that did not have a larger counterpart within the data in order to avoid an estimation bias. This is done by including the scaled open intervals in the estimate.

\begin{equation}
    T_j = (T-t_j)  10^{-b(m_i + \delta m_c)},
\end{equation}
where $t_j$ are times of events of magnitude $m_j$ that are not followed by a larger event (by the margin $\delta m_c$) within the observation interval.

Finally, the a-more-positive a-value estimate is given as
\begin{equation}
a^{++} = \log n^{++} - \log \frac{\sum_{i=1}^{n^{++}} \Delta 
\tau_i + \sum_{j=1}^{m}T_j}{T},
\end{equation}

where $n^{++}$ is the number of closed intervals, $m$ is the number of open intervals, and $n^{++} + m$ is the total number of events above the completeness magnitude. Note that this equation is the same as put forward by \cite{van_der_elst_positive_2023}, with the exception that the total length of the time period $T$ is included in our formulation. This detail has the effect that the classical a-value estimator and the a-more-positive estimator have the same expected value and can therefore be directly compared.

Implementation in SeismoStats: For the estimation of the a-value, the class \texttt{AMorePositiveAValueEstimator} requires the array of magnitudes, the array of times, the magnitude of completeness, the bin size $\Delta m$ and the b-value as input. The user can also specify the threshold $\delta m_c$ as input, with the default value being one magnitude bin. Weights can be given optionally.

\subsection{Pitfalls and Limitations}
Although estimating the parameters of the Gutenberg-Richter (GR) law is straightforward, and the methods implemented in the SeismoStats package have been thoroughly tested, several limitations and potential sources of bias must be considered when applying these methods \citep{herrmann_inconsistencies_2021, marzocchi_how_2020, gulia_contamination_2021}. For example, the $m_c$ estimation methods currently implemented in SeismoStats do not account for spatial or temporal variations in $m_c$, despite evidence that such variability is common in seismic catalogs \citep{schorlemmer_probability_2008}. Catalogs may also contain human-made events, such as quarry blasts \citep{gulia_contamination_2021}. This can introduce artifacts both in the a- and b-value estimation. Furthermore, many catalogs contain different magnitude types, and conversion between magnitude types is not straightforward. Failing to address this heterogeneity can significantly affect the estimated magnitude distribution. \citep[e.g.,][]{deichmann_theoretical_2017}. Users of SeismoStats are therefore encouraged to carefully and critically examine the data before applying statistical methods.

\section{Conclusion}
By simplifying the process of statistical analysis, SeismoStats aims to advance seismological research and contribute to collective efforts to mitigate and prepare for seismic hazards.

Planned future developments include the implementation and improvement of declustering methods, tools for estimating parameters of the Omori-Utsu law \citep{utsu_centenary_1995}, calculating the fractal dimension of seismicity \citep{kagan_earthquake_2007}, and functionalities to rapidly compare different earthquake catalogs. Furthermore, our objective is to improve interoperability and compatibility with other established seismological packages such as pyCSEP \citep{savran_pycsep_2022, graham_new_2024} to support collaborative development across the seismological community.

We strongly encourage the community to contribute to the ongoing enhancement of SeismoStats. Researchers and developers are invited to share suggestions, propose new features, or contribute code through our \href{https://github.com/swiss-seismological-service/SeismoStats}{GitHub repository}. Through this collaborative effort, we hope to advance SeismoStats and support progress in seismological research.

\section{Data and Resources}
All figures provided in this article are based on the Swiss catalog between 1 January 2024 and 1 January 2025 \citep{swiss_seismological_service_national_1983}.

The complete documentation can be found here: \url{https://seismostats.readthedocs.io/latest/}. The corresponding code repository can be found here: \url{https://github.com/swiss-seismological-service/SeismoStats}

\section{Declaration of Competing Interests}
The authors declare no competing interests.

\section{Acknowledgments}
We thank Pablo Iturrieta and Chris Rollins for many valuable comments. This research is funded by CETPartnership, the Clean Energy Transition Partnership under the 2023 joint call for research proposals project "GEOWTINS", co-funded by the European Commission (GA N°101069750) and the Swiss Federal Office of Energy (SI/502835-01).

\end{document}